\documentclass[twoside,twocolumn,9pt]{article}
\usepackage{extsizes}
\usepackage[super,sort&compress,comma]{natbib} 
\usepackage[version=3]{mhchem}
\usepackage[left=1.5cm, right=1.5cm, top=1.785cm, bottom=2.0cm]{geometry}
\usepackage{balance}
\usepackage{mathptmx}
\usepackage{sectsty}
\usepackage{graphicx} 
\usepackage{lastpage}
\usepackage[format=plain,justification=justified,singlelinecheck=false,font={stretch=1.125,small,sf},labelfont=bf,labelsep=space]{caption}
\usepackage{float}
\usepackage{fancyhdr}
\usepackage{fnpos}
\usepackage[english]{babel}
\addto{\captionsenglish}{%
  
}
\usepackage{array}
\usepackage{droidsans}
\usepackage{charter}
\usepackage[T1]{fontenc}
\usepackage[usenames,dvipsnames]{xcolor}
\usepackage{setspace}
\usepackage[compact]{titlesec}
\usepackage{hyperref}
\usepackage{chemformula}

\usepackage{epstopdf}

\definecolor{cream}{RGB}{222,217,201}

\begin{document}

\pagestyle{fancy}
\thispagestyle{plain}
\fancypagestyle{plain}{
\renewcommand{\headrulewidth}{0pt}
}

\makeFNbottom
\makeatletter
\renewcommand\LARGE{\@setfontsize\LARGE{15pt}{17}}
\renewcommand\Large{\@setfontsize\Large{12pt}{14}}
\renewcommand\large{\@setfontsize\large{10pt}{12}}
\renewcommand\footnotesize{\@setfontsize\footnotesize{7pt}{10}}
\makeatother

\renewcommand{\thefootnote}{\fnsymbol{footnote}}
\renewcommand\footnoterule{\vspace*{1pt}%
\color{cream}\hrule width 3.5in height 0.4pt \color{black}\vspace*{5pt}} 
\setcounter{secnumdepth}{5}

\makeatletter 
\renewcommand\@biblabel[1]{#1}            
\renewcommand\@makefntext[1]%
{\noindent\makebox[0pt][r]{\@thefnmark\,}#1}
\makeatother 
\renewcommand{\figurename}{\small{Fig.}~}
\sectionfont{\sffamily\Large}
\subsectionfont{\normalsize}
\subsubsectionfont{\bf}
\setstretch{1.125} 
\setlength{\skip\footins}{0.8cm}
\setlength{\footnotesep}{0.25cm}
\setlength{\jot}{10pt}
\titlespacing*{\section}{0pt}{4pt}{4pt}
\titlespacing*{\subsection}{0pt}{15pt}{1pt}

\fancyfoot{}
\fancyfoot[LO,RE]{\vspace{-7.1pt}\includegraphics[height=9pt]{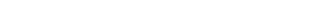}}
\fancyfoot[CO]{\vspace{-7.1pt}\hspace{13.2cm}\includegraphics{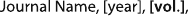}}
\fancyfoot[CE]{\vspace{-7.2pt}\hspace{-14.2cm}\includegraphics{RF}}
\fancyfoot[RO]{\footnotesize{\sffamily{1--\pageref{LastPage} ~\textbar  \hspace{2pt}\thepage}}}
\fancyfoot[LE]{\footnotesize{\sffamily{\thepage~\textbar\hspace{3.45cm} 1--\pageref{LastPage}}}}
\fancyhead{}
\renewcommand{\headrulewidth}{0pt} 
\renewcommand{\footrulewidth}{0pt}
\setlength{\arrayrulewidth}{1pt}
\setlength{\columnsep}{6.5mm}
\setlength\bibsep{1pt}

\makeatletter 
\newlength{\figrulesep} 
\setlength{\figrulesep}{0.5\textfloatsep} 

\newcommand{\topfigrule}{\vspace*{-1pt}%
\noindent{\color{cream}\rule[-\figrulesep]{\columnwidth}{1.5pt}} }

\newcommand{\botfigrule}{\vspace*{-2pt}%
\noindent{\color{cream}\rule[\figrulesep]{\columnwidth}{1.5pt}} }

\newcommand{\dblfigrule}{\vspace*{-1pt}%
\noindent{\color{cream}\rule[-\figrulesep]{\textwidth}{1.5pt}} }

\makeatother

\twocolumn[
  \begin{@twocolumnfalse}
{\includegraphics[height=30pt]{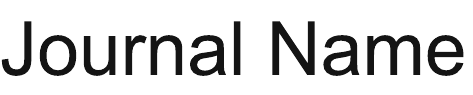}\hfill\raisebox{0pt}[0pt][0pt]{\includegraphics[height=55pt]{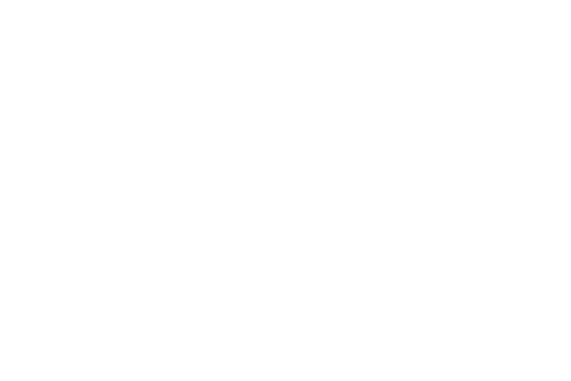}}\\[1ex]
\includegraphics[width=18.5cm]{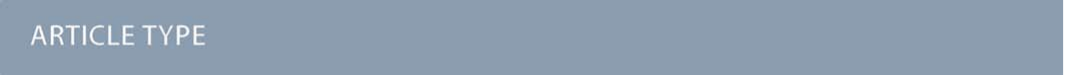}}\par
\vspace{1em}
\sffamily
\begin{tabular}{m{4.5cm} p{13.5cm} }

\includegraphics{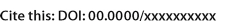} & \noindent\LARGE{\textbf{X-raying Mg$_{0.2}$Co$_{0.2}$Ni$_{0.2}$Cu$_{0.2}$Zn$_{0.2}$O: disentangling elemental contributions in a prototypical high-entropy oxide}} \\
\vspace{0.3cm} & \vspace{0.3cm} \\

 & \noindent\large{Maryia Zinouyeva,$^{\ast}$\textit{$^{a}$} Martina Fracchia,\textit{$^{b,c}$} Giulia Maranini,\textit{$^{b}$}  Mauro Coduri,\textit{$^{b,c}$} Davide Impelluso,\textit{$^{a}$}  Nicholas B. Brookes,\textit{$^{d}$} Lorenzo Grilli,\textit{$^{a\dag}$} Kurt Kummer,\textit{$^{d}$} Francesco Rosa,\textit{$^{a}$} Matteo Aramini,\textit{$^{e}$} 
 Giacomo Ghiringhelli, \textit{$^{a,f}$} Paolo Ghigna,\textit{$^{b,c}$} and Marco  Moretti Sala.\textit{$^{a\ddag}$}} \\

\includegraphics{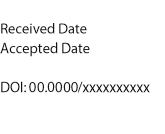} & \noindent\normalsize{We employ several X-ray based techniques, including X-ray diffraction, absorption and resonant inelastic scattering, to disentangle the contributions of individual chemical species to the structural, electronic and magnetic properties of high-entropy oxides. In the benchmark compound Mg$_{0.2}$Co$_{0.2}$Ni$_{0.2}$Cu$_{0.2}$Zn$_{0.2}$O and related systems, we unambiguously resolve a sizable Jahn-Teller distortion at the Cu sites, more pronounced in the absence of \ch{Ni^{2+}} and \ch{Mg^{2+}}, suggesting that these ions promote positional order, whereas \ch{Cu^{2+}} ions act to destabilize it. Moreover, we detect magnetic excitations and estimate the strength of the interactions between pairs of different magnetic elements. Our results provide valuable insights into the role of the various chemical species in shaping the physical properties of high-entropy oxides.} \\

\end{tabular}

 \end{@twocolumnfalse} \vspace{0.6cm}

  ]

\renewcommand*\rmdefault{bch}\normalfont\upshape
\rmfamily
\section*{}
\vspace{-1cm}


\footnotetext{\textit{$^{a}$~Dipartimento di Fisica, Politecnico di Milano, piazza Leonardo da Vinci 32, I-20133 Milano, Italy}}
\footnotetext{\textit{$^{b}$~Department of Chemistry, University of Pavia, V.le Taramelli, 12, I-27100 Pavia, Italy.}}
\footnotetext{\textit{$^{c}$~INSTM, National Inter-University Consortium for Materials Science and Technology, Via G. Giusti 9, I-50121 Florence, Italy.}}
\footnotetext{\textit{$^{d}$~ESRF, The European Synchrotron, 71 Avenue des Martyrs, CS 40220, F-38043 Grenoble, France}}
\footnotetext{\textit{$^{e}$~Diamond Light Source, Harwell Science and Innovation Campus, Didcot, OX11 0DE, United Kingdom.}}
\footnotetext{\textit{$^{f}$~CNR-SPIN, Dipartimento di Fisica, Politecnico di Milano, I-20133 Milano, Italy}}
\footnotetext{\textit{$^{\ast}$~e-mail: maryia.zinouyeva@polimi.it}}
\footnotetext{\textit{$^{\dag}$~current address: Institut für Experimentalphysik, Freie Universität Berlin, Arnimallee 14, D-14195 Berlin-Dahlem, Germany}}
\footnotetext{\textit{$^{\ddag}$~e-mail: marco.moretti@polimi.it}}





\section{\label{sec:Introduction}Introduction}

High-entropy materials are single-phase compounds characterized by significant configurational disorder, wherein multiple chemical species randomly occupy equivalent crystallographic sites. Configurational disorder disrupts periodicity and symmetry, significantly influencing a material's physical properties. Among high-entropy materials, high-entropy oxides (HEOs) stand out for their exceptional characteristics\cite{FRACCHIA2024585}, such as tunable room-temperature superionic conductivity \cite{berardan2016}, colossal dielectric constants \cite{berardan2016colossal}, and high lithium-storage capacity \cite{sarkar2018, qiu2019}.
In HEOs configurational disorder facilitates the formation of a single-phase material from the multi-phase mixture of different chemical species through a temperature-dependent entropic contribution $-T\Delta S_\mathrm{c}<0$, which offsets the enthalpic contributions $\Delta H_\mathrm{mix}$ to the Gibbs free energy of mixing $\Delta G_\mathrm{mix}=\Delta H_\mathrm{mix} - T \Delta S_\mathrm{c}$; specifically, the spontaneous formation of a homogeneous solid solution occurs when the latter is negative. So, for entropy-stabilized oxides (ESOs), i.e., HEOs where the enthalpic contribution is positive and the entropic contribution satisfies $T \Delta S_\mathrm{c}>\Delta H_\mathrm{mix}$, there exists a critical temperature $T_c = \Delta H_\mathrm{mix}/\Delta S_\mathrm{c}$ below which the homogeneous solid solution cannot spontaneously form. Since $\Delta S_\mathrm{c} = -R \sum_{i=1}^{N} \chi_i \ln{\chi_i}$, where $\chi_i$ is the molar fraction of the $i$-th chemical species and $N$ is their total number, configurational entropy is maximized in equimolar systems, for which $\chi_i = 1/N$ and $\Delta S_\mathrm{c} = R \ln{N}$.

Rocksalt Mg$_{0.2}$Ni$_{0.2}$Co$_{0.2}$Zn$_{0.2}$Cu$_{0.2}$O (hereafter referred to as 5HEO) is the prototypical HEO\cite{rost2015}. It is obtained from an equimolar mixture of MgO, CoO, NiO, CuO and ZnO, which, however, do not share a common crystal structure; indeed, MgO, CoO and NiO possess rocksalt, while CuO and ZnO tenorite and wurtzite crystal structures, respectively, such that their incorporation into a rocksalt structure comes with a positive enthalpy of mixing. Rost \emph{et al.} \cite{rost2015} showed that i) a single-phase crystal of 5HEO forms only for an equilibration temperature $T_c\approx 1150$\,K or above; since $\Delta S_\mathrm{c}=13.4$\,J/(mol\,K) for $N=5$, such a critical temperature corresponds to an enthalpy of mixing of approximately 15.4\,kJ/mol; ii) the transition from the multi- to the single-phase solid solution is reversible with temperature; iii) the critical temperature is minimized for the equimolar composition when the amount of an individual chemical species is changed and the others are kept even, and iv) the quaternary compounds never form a single phase at $T_c$ or below. Moreover, v) the authors' estimate of 10\,kJ/mol for the enthalpy of mixing solely based on the tenorite-to-rocksalt (22.2\,kJ/mol\cite{Bularzik1986}) and the wurtzite-to-rocksalt (24.5\,kJ/mol\cite{Davies1981}) transition enthalpies for CuO and ZnO, respectively, is compatible with the experimental estimate in iii). These observations were taken as evidence that the formation of the homogeneous solid solution is driven by entropy, i.e., that 5HEO is an ESO. However, iii) suggests that the enthalpy of mixing is very little dependent on the actual composition of the material, implying that the enthalpy of mixing of all the chemical species is very similar. This is in contradiction with the hypothesis used in v), where the enthalpy of mixing of MgO, CoO and NiO were neglected. Indeed, Coduri \emph{et al}. find that the relative concentration of MgO, CoO and NiO affects the stability of the rocksalt structure in samples with the same configurational entropy, suggesting that also the nature of the chemical species with rocksalt structure also plays a role \cite{Coduri2024}. A similar conclusion can be drawn from the observation of Fracchia \emph{et al.} that \ch{Ni_{0.6}Cu_{0.2}Zn_{0.2}O} is stable in the rocksalt structure at relatively low temperatures \cite{fracchia2022, FRACCHIA2025117237}; here, configurational entropy is lower, so that the critical temperature should be higher than 5HEO if CuO and ZnO were the sole chemical species contributing to the enthalpy of mixing. The most plausible scenario is that the latter hypothesis should be relaxed. Recently, experimental \cite{berardan2017, rost2017, fracchia2020, sushil2021} and theoretical \cite{rak2016, rak2018, anand2018, kaufman2021} studies have provided evidence for Jahn-Teller distortions at the sites occupied by \ch{Cu} in 5HEO. Their very presence points at deviations of the crystal structure from the perfect rocksalt, which should definitely be taken into account for a quantitative understanding of the energetics of the system. Indeed, distortions of the sites occupied by \ch{Cu} are necessarily shared with neighboring sites, which are randomly occupied by various chemical elements that may respond very differently to the perturbation. Interestingly, this could explain the dependence of the stability of the rocksalt structure on the relative concentration of the various chemical species; unfortunately,  structural probes did not reveal distortions of the octahedral environment around any of the metal ions, with the exception of \ch{Cu^{2+}}. Such a picture is complicated by the tendency of 5HEO to form the guggenite phase rather than simply demixing into the single binary oxides below the critical temperature \cite{Coduri2025}. 
All the above findings prompted us to adopt a different approach, focusing on potential spectroscopic signatures of the putative distortions on the electronic structure of the various metal ions, individually.

Another intriguing feature of 5HEO is that, despite configurational disorder, it displays long-range magnetic order below approximately 113\,K \cite{zhang2019}. Indeed, neutron powder diffraction studies revealed a magnetic structure analogous to that of NiO and CoO \cite{roth1958}, with antiferromagnetically coupled ferromagnetic (111) planes \cite{zhang2019}. Notably, spin waves have been observed at temperatures significantly higher than the Néel temperature, suggesting the persistence of short-range magnetic correlations also in the paramagnetic state \cite{zhang2019}. Magnetic susceptibility measurements \cite{jimenez2019} and Monte Carlo simulations \cite{rak2020} indicate that the long-range magnetic order is strongly dependent on the concentration of magnetic ions, again pointing at the nature of the chemical species playing a role in shaping the HEO's physical properties. Unfortunately, the contribution of the individual metal ions to the collective magnetic behavior could not be singled out, another challenge that we address in this paper.

We performed X-ray diffraction (XRD), including pair distribution function (PDF) analysis, as well as soft- and hard-X-ray absorption spectroscopy (XAS) and resonant inelastic X-ray scattering (RIXS) measurements in 5HEO and in five sister compounds. RIXS is a powerful probe of the electronic and magnetic excitations in solids and benefits from its intrinsic chemical sensitivity, which is particularly welcome in the context of HEOs to isolate the contributions of the various metal ions. Our results directly evidence a sizable distortion of the \ch{Cu^{2+}} sites and provide hints of the associated energy scale. Moreover, unlike structural probes, RIXS provides indications that the sites occupied by \ch{Co} and \ch{Ni} are also distorted, and we point at the different roles played by the various metal ions in accommodating these distortions. Finally, we determine almost all of the relevant magnetic interactions among pairs of magnetic ions, which are essential ingredients for a quantitative understanding of the magnetic properties of HEOs.

\section{Experimental method}\label{sec:Method}
The six powder samples studied in this work are listed in Tab.\,\ref{tab:Composition}, together with their crystal structure and cell parameter. 

\begin{table*}
\centering
\renewcommand{\arraystretch}{1.5}
\begin{tabular}{ccccc}
 \hline
 Chemical composition &  label  & structure & cell parameter(s) (\AA) & FWHM ratio \\
 \hline
   \ch{Mg_{0.25}Co_{0.25}Ni_{0.25}Zn_{0.25}O} & noCu & cubic & $a=4.2301\pm0.00012$ & 1.0 \\
\ch{Mg_{0.20}Co_{0.20}Ni_{0.20}Cu_{0.20}Zn_{0.20}O} & 5HEO & cubic & $a=4.2365\pm0.00006$ & 1.1 \\
 \ch{Mg_{0.25}Ni_{0.25}Cu_{0.25}Zn_{0.25}O} & noCo & cubic & $a=4.2293\pm0.00008$ & 1.3 \\
\ch{Mg_{0.25}Co_{0.25}Ni_{0.25}Cu_{0.25}O} & noZn & cubic & $a=4.2258\pm0.00018$ & 1.8 \\
 \ch{Co_{0.25}Ni_{0.25}Cu_{0.25}Zn_{0.25}O} & noMg & tetragonal & $a=3.0138\pm0.00052$, $c=4.2106\pm0.00097$ & \\
  \ch{Mg_{0.25}Co_{0.25}Cu_{0.25}Zn_{0.25}O} & noNi & tetragonal & $a=3.0311\pm0.00039$, $c=4.1936\pm0.00062$ & \\
\hline
\end{tabular}
\caption{Chemical composition, label, crystal structure, and lattice parameter(s) of the samples. For the cubic systems, the ratio of the FWHM between the 200 and 111 reflections is also reported.}\label{tab:Composition}
\end{table*}

The samples were prepared using the Pechini method, a variant of the sol-gel method. A starting solution was obtained by dissolving a stoichiometric amount of nitrate salts of each cation in distilled water, with an excess of citric acid. The solution was stirred and heated at $100^\circ{}$C overnight to form a gel which was subsequently treated at $280^\circ{}$C for two hours and then at $900^\circ{}$C for another two hours to remove the organic components. The resulting fine powders consisted primarily of mechanically blended oxide precursors. These powders were pelletized and annealed at $1000^\circ{}$C for 12 hours to complete the solid solution formation and then quenched at room temperature. 

XRD patterns were recorded in Bragg-Brentano geometry using a Bruker D2 diffractometer equipped with Cu radiation and a Ni filter collecting data over a 2$\theta$ angular range from $15^\circ{}$ to $90^\circ{}$. Rietveld refinements were performed with GsasII software to verify the crystal structure and determine the lattice parameters \cite{Toby2013}. The full-width-at-half-maximum (FWHM) has been computed using the software WINPlotR \cite{Roisnel2001}. XRD patterns for PDF analysis were collected at beamline ID15A of ESRF – The European Synchrotron (Grenoble, France) at an incident wavelength of $\lambda$ = 0.1240\,\AA\ ($\approx 100$\,keV), using a Pilatus 2M CdTe detector (Dectris) placed 380\,mm away from the specimen \cite{Vaughan2020}. The powders were packed into quartz capillaries (Hilgenberg) with 0.5\,mm diameter and rotated during the acquisition to improve the orientational average of the grains. PDF data were processed using the software PDFGetX3 \cite{Juhas2013} limiting the integration of diffraction data to a maximum value of the momentum transfer of 24\,\AA$^{-1}$.

Hard XAS at the \ch{Co}, \ch{Ni}, and \ch{Cu} K edges were carried out at beamline I20 of Diamond Light Source (Didcot, United Kingdom) on 5HEO, and at beamline XAFS of ELETTRA (Trieste, Italy) on all the other samples. In both cases, the spectra were acquired at room temperature.
The X-ray absorption near edge spectroscopy (XANES) part of the spectra was analyzed by means of the Athena code \cite{Ravel2005}. Spectra were first processed by subtracting a smooth pre-edge background fitted with a straight line, and then normalized to unity at 300 eV above the edge jump, where the extended X-ray absorption fine structure (EXAFS) can be neglected. The EXAFS fittings were performed using the EXCURVE program \cite{Feiters2020-qx}. 

Soft XAS and RIXS measurements were carried out at beamline ID32 of ESRF – The European Synchrotron (Grenoble, France) \cite{BROOKES2018175}. The spectra were collected at the \ch{Co}, \ch{Ni}, and \ch{Cu} L$_3$ edges with energy resolutions of approximately 40, 46, and 50\,meV, respectively, using scattering angles of $90^\circ{}$ and $150^\circ{}$ with $\pi$ polarization of the incident photons. Measurements were performed at 30, 300, and 600\,K. The RIXS spectra were normalized to the integrated intensity of the crystal-field excitations, specifically in the energy range from 0.5 to 3.5\,eV at the \ch{Ni} and \ch{Co} L$_3$ edges and from 0.5 to 2\,eV at the \ch{Cu} L$_3$ edge.

\section{Results and discussion}\label{sec:Discussion}
\begin{figure}
   \includegraphics[width=\columnwidth]{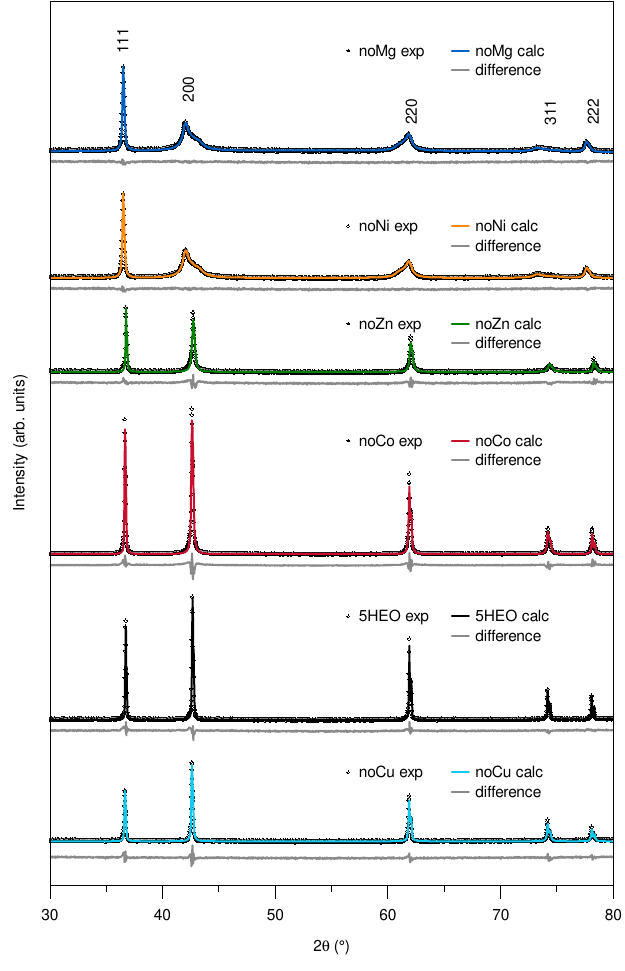}
    \caption{Rietveld refinements of XRD patterns collected at room temperature.}
       \label{fig:XRD}
\end{figure}

\subsection{Long-range crystal structure -- XRD}
To start with, we characterize the crystal structure of the investigated samples by means of XRD. The corresponding patterns are displayed in Fig.\,\ref{fig:XRD}, while the results of Rietveld refinements are summarized in Tab.\,\ref{tab:Composition}. Four samples (5HEO, noCu, noCo, and noZn) are consistent with the prototype rocksalt crystal structure (space group $Fm\bar{3}m$). Among them, noCu and 5HEO display sharp peaks, whereas noCo and noZn show a general peak broadening of all reflections except for those related to the $hhh$ family \cite{BERARDAN2017693}. Such broadening is then quantified as the ratio between the full-width-at-half-maximum (FWHM) of the 200 and 111 reflections (see Tab.\,\ref{tab:Composition}): this is the smallest for noCu and increases in 5HEO, noCo and noZn, yet without inducing a structural transition. A transition from the cubic to the tetragonal crystal structure (space group $I4/mmm$), instead, occurs in noMg and noNi, as evidenced by the splitting and asymmetric broadening of all, but the peaks corresponding to the $hhh$ reflections. Our results are in line with available literature  \cite{Coduri2025,D1TC03287A}. 

\subsection{Short-range crystal structure -- PDF}
\begin{figure*}
    \centering
   \includegraphics[width=0.75\textwidth]{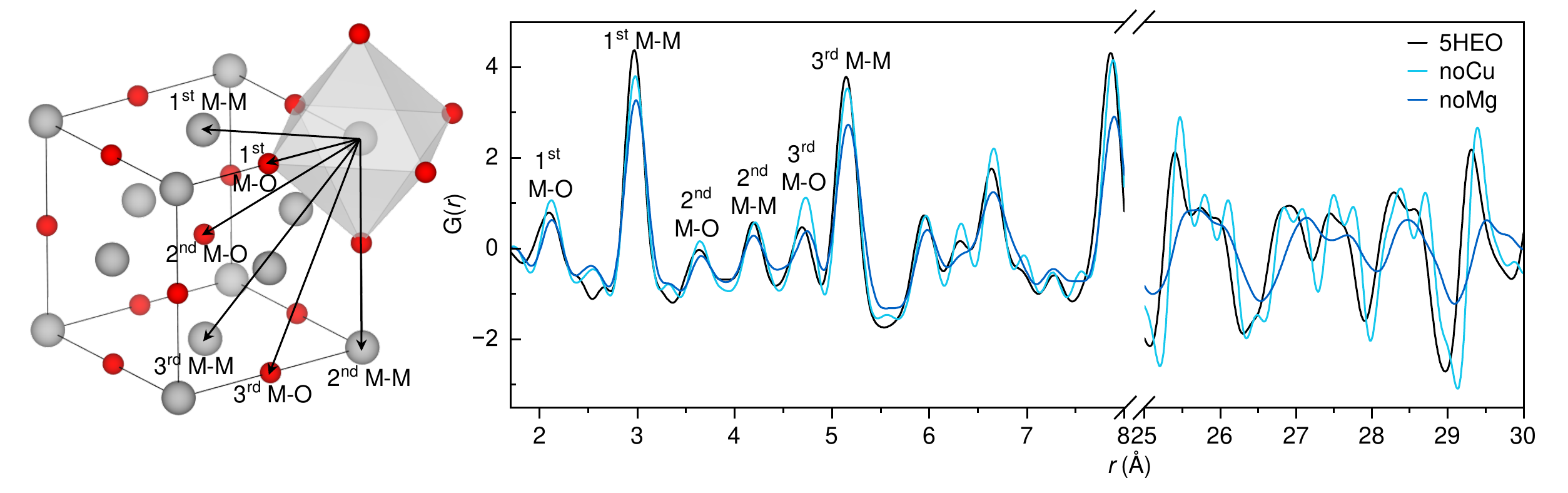}
    \caption{Sketch of the rocksalt unit cell, with grey and red spheres representing the metal  and the oxygen ions, respectively, and the PDF curves of 5HEO, noCu and noMg. The principal interatomic distances used to interpret the PDF data, as well as a \ch{MO_6} octahedron are shown. }
       \label{fig:PDF}
\end{figure*}


\begin{table}
\centering
\renewcommand{\arraystretch}{1.5}
\begin{tabular}{ccccccc}
 \hline
 atom pair & distance (\AA) & length & multiplicity \\  
 \hline
 1\textsuperscript{st} M-O & $2.1183\pm0.00003$ & $a/2$ & 6\\
 1\textsuperscript{st} M-M & $2.9957\pm0.00004$ & $a\sqrt{2}/2$ & 12\\
 2\textsuperscript{nd} M-O & $3.6689\pm0.00005$ & $a\sqrt{3}/2$ & 8\\
 2\textsuperscript{nd} M-M & $4.2365\pm0.00006$ & $a$ & 6\\
 3\textsuperscript{rd} M-O & $4.7366\pm0.00007$ & $\sqrt{\left(a/2\right)^2+a^2}$ & 24\\
 3\textsuperscript{rd} M-O & $5.1886\pm0.00007$ & $\sqrt{\left(a\sqrt{2}/2\right)^2+a^2}$  & 24\\
 \hline
\end{tabular}
\caption{Labeling of the first six peaks observed in the $G(r)$ curves of Fig.\,\ref{fig:PDF}: nature of the involved species, interatomic distance, dependence on the cell parameter and multiplicity.}\label{tab:pdf}
\end{table}

The broadening observed in XRD patterns suggest a certain degree of long-range disorder, eventually resulting in a structural transition in some of the samples. Insights on the short-range, i.e., local, disorder can, instead, be obtained by performing a PDF analysis. Fig.\,\ref{fig:PDF} shows the experimental PDF curves in $G(r)$ notation for 5HEO, noCu and noMg, where the latter two have been chosen as references for cubic and tetragonal crystal structures, respectively. The atom pairs associated to the PDF peaks are identified in Tab.\,\ref{tab:pdf}, along with the corresponding interatomic distances for the rocksalt phase and their estimate for $a = 4.2365$\,\AA, the cell parameter of 5HEO. We assume that O-O pairs do not contribute to the PDF peaks due to the low scattering factor of oxygen compared to metal ions. 
The PDF curves in Fig.\,\ref{fig:PDF} of noCu, 5HEO and noMg are similar, but with appreciable differences, particularly in terms of peak position and broadening at large interatomic distances. Since the broadening of the peaks mostly depends on the degree of local disorder, we conclude that noCu is more ordered compared to 5HEO and, notably, noMg, which appears to be the most disordered, in agreement with previous high-resolution powder diffraction measurements \cite{Coduri2025}. The fact that the PDF signal of tetragonal noMg is similar to that of cubic noCu and 5HEO at short interatomic distances is not surprising because the PDF analysis is limited by thermal vibration and termination ripples. Rather, the impact of the tetragonal distortion is evident at large interatomic distances (above 20\,\AA), where the curve of noMg is markedly different from the others.

\subsection{Short-range crystal structure -- XANES and EXAFS}
\begin{figure*}
    \centering
   \includegraphics[width=0.75\textwidth]{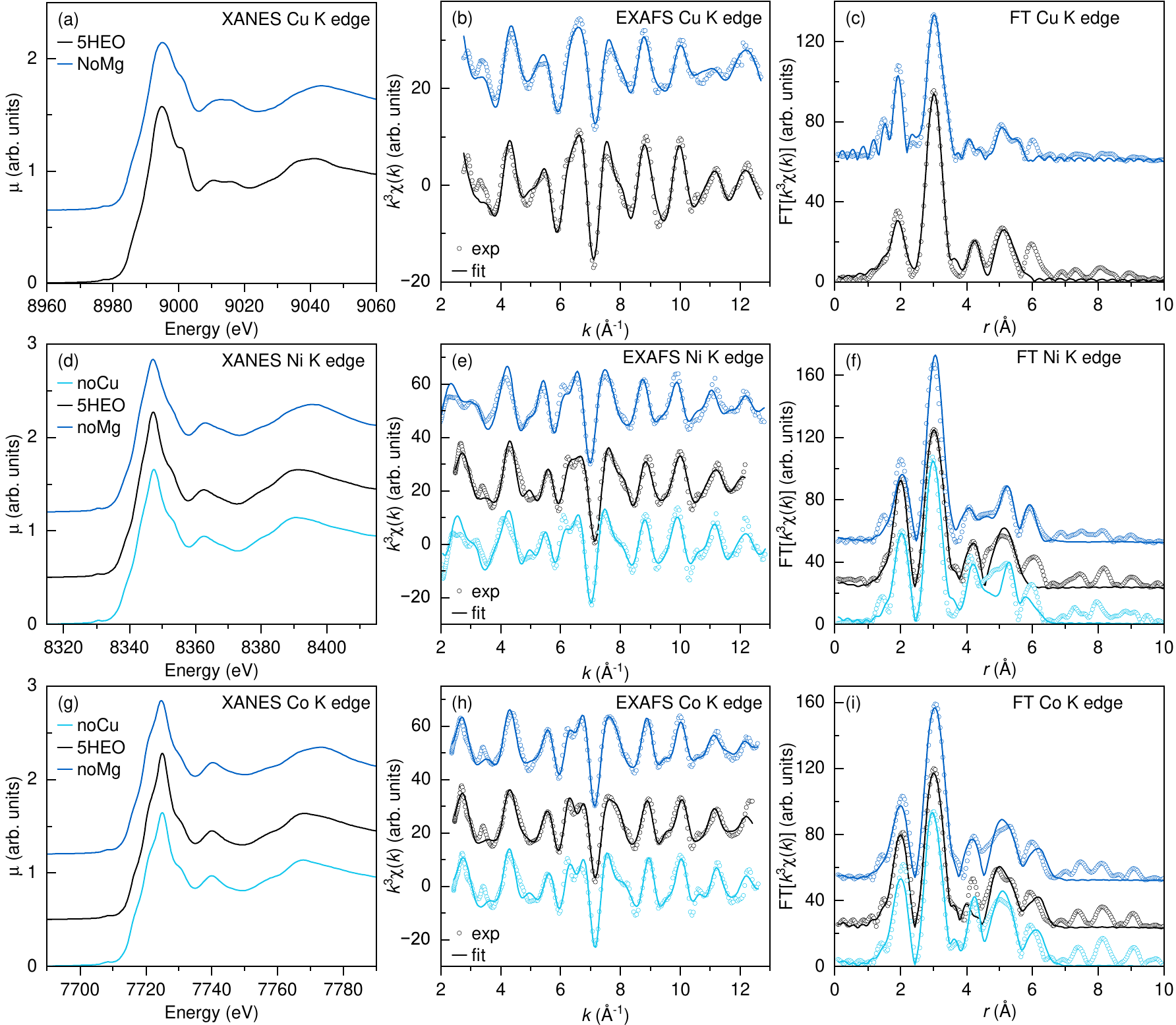}
    \caption{Cu (a), Ni (b) and Co (c) K edge XANES spectra for noCu, 5HEO and noMg samples. The corresponding experimental (open circles) and simulated (thin line) EXAFS oscillations are reported in panels (b), (e) and (h), and (c), (f) and (i), respectively.}
       \label{fig:XAS}
\end{figure*}
\begin{figure}
   \includegraphics[width=0.45\textwidth]{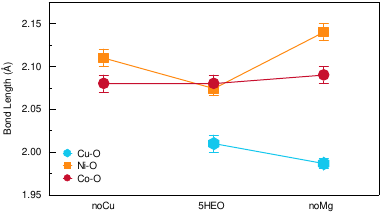}
    \caption{Average Co-O, Ni-O and Cu-O bond lengths for noCu, 5HEO and noMg  as extracted from Cu, Ni and Co K edge EXAFS, respectively.}
       \label{fig:bond-lengths}
\end{figure}

Structural characterization techniques evidence a certain degree of long- and short-range disorder in some of the samples. Unfortunately, the information they provide is averaged over all metal sites. With the aim to characterize the short-range crystal structure around the various chemical species individually, we acquired hard XAS spectra at the Co, Ni and Cu K edges for noCu, 5HEO and noMg in Fig.\,\ref{fig:XAS}. The XANES data for the Cu, Co and Ni K edges are reported in panels (a), (d) and (g), respectively. We note that spectra features are progressively less resolved in the XANES spectra of 5HEO and, particularly, noMg than in the noCu sample. This trend is also evident in the EXAFS spectra, shown in panels (b), (e) and (h), along with the corresponding Fourier transforms (FT) in panels (c), (f) and (i). To make the above discussion more quantitative, the EXAFS oscillations and their FT were fitted adopting a rocksalt structure for all the samples, even though noMg possesses a tetragonal crystal structure. The reason behind this choice is twofold: on one hand, we keep the number of fitting parameters as low as possible, in order to avoid unnecessary correlations between them; on the other hand, the change in the bond lengths due to the tetragonal distortion is 0.03\,\AA\ at the most, i.e., comparable with the experimental resolution of EXAFS, and, as a matter of fact, the first peak in the FT of noMg is never split. The fit to the EXAFS oscillations and their FT are shown by the thin lines in Figs.\,\ref{fig:XAS} and the extracted average Co-O, Ni-O and Cu-O bond lengths are shown in Fig.\,\ref{fig:bond-lengths}. We note that the Co-O bond length remains practically unchanged and identical within the errorbars to the Ni-O bond length in 5HEO, while the Cu-O and Ni-O bond lengths undergo opposite changes, with the former decreasing and the latter increasing in the tetragonal crystal structure. We interpret this and the above observations as an effect of Jahn-Teller distortions at the Cu sites, which induce a compression of the axial ligands, resulting in a smaller average Cu-O distance\cite{Coduri2025}. It is peculiar, though, that the change in the Cu-O bond length has no or little impact on the Co-O, but strong on the Ni-O bond length, indicating that Ni exhibits a greater propensity than Co to adjust to the distortions introduced by Cu$^{2+}$. Indeed, in a regular octahedral environment, the Ni-O atom pair should be shorter than Co-O and Cu-O, according to the Shannon's ionic radii\cite{Shannon1976}.

\subsection{\label{subsec:Cu L3_dd} Crystal-field excitations and Jahn-Teller distortions}

The interpretation of XRD and hard XAS data based on the Jahn-Teller distortion at the Cu sites is rather speculative at this stage and necessitates further investigation. RIXS can be extremely helpful in this respect because it directly probes $d$-$d$ excitations. For a transition metal ion in octahedral symmetry, the crystal field splits the $d$ orbitals into an $e_g$ ($d_{x^2-y^2}$ and $d_{3z^2-r^2}$) doublet and a $t_{2g}$ ($d_{xy}$, $d_{yz}$ and $d_{zx}$) triplet; for lower symmetries, these degeneracies are further lifted. So-called $d$-$d$ excitations are transitions between crystal-field split states and their energy is therefore directly linked to the strength and symmetry of the crystal field interaction \cite{sala2011}. 

\begin{figure*}
    \centering
   \includegraphics[width=0.75\textwidth]{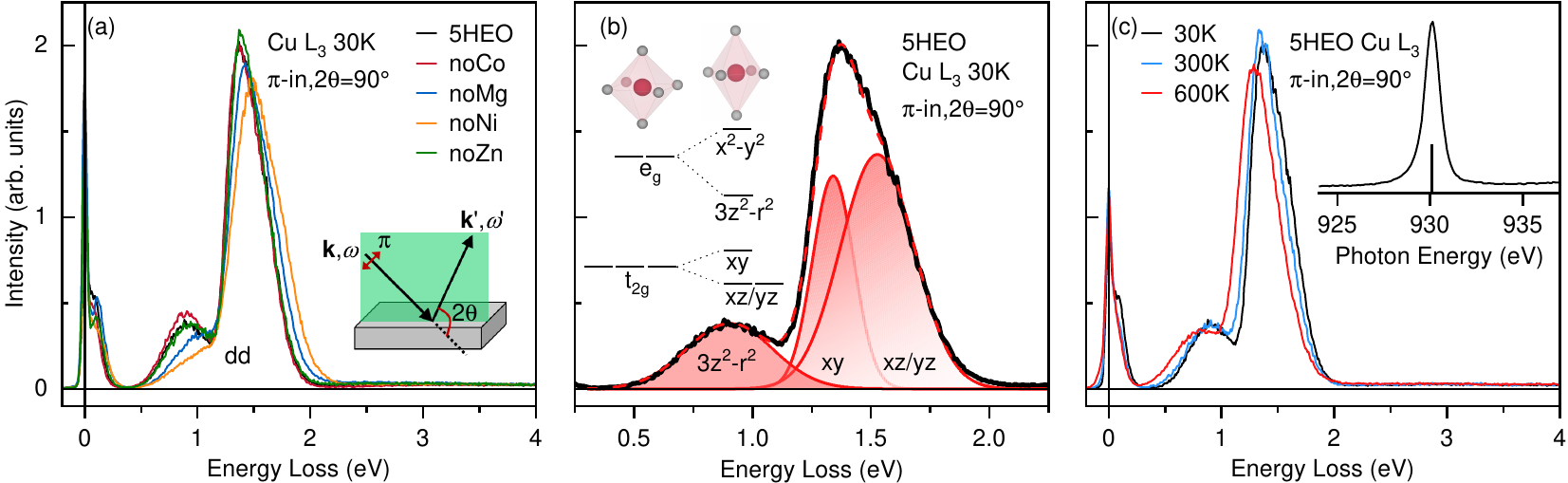}
    \caption{(a) \ch{Cu} L$_3$ edge RIXS spectra taken at 30\,K with an energy of the incident photons corresponding to the main peak in the XAS profile (vertical red bar in the inset of panel (c)). The  scheme of the experimental RIXS geometry is shown in the inset. (b) Experimental and fitting curves of the $d$-$d$ excitations in 5HEO at 30\,K as explained in the text. The inset shows the crystal field splitting of $e_g$ and $t_{2g}$ levels in a Jahn-Teller distorted octahedral system. (c) Temperature dependence of the RIXS spectra of 5HEO.}
       \label{fig:Cuedge}
\end{figure*}

Fig.\,\ref{fig:Cuedge}\,(a) shows the collected RIXS spectra at the \ch{Cu} L$_3$ edge (approximately 930\,eV, see inset in panel (c)) at 30\,K for the five samples containing \ch{Cu}. In analogy to the literature of insulating and superconducting cuprates \cite{GhiringhelliPRL2004, sala2011, oles2019, Barantani_PRX}, the peaks in the energy range from 0.5 to 2\,eV are assigned to $d$-$d$ excitations and the weak, but extended spectral weight above 2\,eV is attributed to charge-transfer excitations. Similarly, we could tentatively assign the peak at approximately 0.1\,eV to magnetic excitations, albeit the lack of long-range configurational order necessitates a degree of caution. In this section, we focus on $d$-$d$ excitations, while magnetic excitations will be discussed in the following.\par
In \ch{Cu^{2+}} with a 3$d^9$ electronic configuration a single hole occupies the $d_{x^2-y^2}$ orbital. The RIXS cross-section for $d$-$d$ excitations in \ch{Cu^{2+}} is well described by a single-ion model \cite{sala2011}, which is here exploited to assign features in the RIXS spectra based on their intensities. To do so, we fit the region of the spectra between 0.5 and 2\,eV to three Gaussian curves whose integrated spectral weight is constrained to the calculated one using the actual experimental conditions and performing the powder average, see Fig.\,\ref{fig:Cuedge}\,(b).


\begin{table*}[]
\centering
\renewcommand{\arraystretch}{1.5}
\begin{tabular}{cccccc}
 \hline
  &  $d_{3z^2-r^2}$ & $d_{xy}$ & $d_{xz/yz}$ & $\Delta E^{\mathrm{Cu}}$ & $\Delta E^{\mathrm{Ni}}$\\
 \hline
 5HEO & $0.904\pm 0.007$ ($0.430\pm0.014$)&$1.341\pm 0.002$ ($0.205\pm0.004$)&$1.527\pm 0.004$ ($0.366\pm0.006$)&  $90.5\pm6.6$ & $43.5\pm6.4$\\
 noCo & $0.888\pm0.005$ ($0.374\pm0.012$)&$1.338\pm 0.002$ ($0.208\pm0.004$)&$1.517\pm 0.004$ ($0.375\pm 0.006$)&  $86.7\pm6.8$ & $44.5\pm2.7$\\
 noCu &   & & &  &$44.4\pm2.6$\\
 noMg & $1.023\pm 0.011$ ($0.550\pm 0.020$)&$1.389\pm 0.003$ ($0.240\pm 0.004$)&$1.580\pm 0.004$ ($0.415\pm 0.007$)&  $114.8 \pm 2.7$ &$ 52.1\pm 4.0$\\
 noNi & $1.173\pm 0.017$ ($0.760\pm 0.044$)&$1.431\pm 0.004$ ($0.281\pm 0.006$)&$1.637\pm 0.005$ ($0.456\pm0.009$) &$96.4 \pm 16.4$ &\\ 
 noZn & $0.920\pm 0.007$ ($0.434\pm0.014$)&$1.362\pm 0.002$ ($0.210\pm 0.004$)&$1.529\pm 0.004$ ($0.387\pm 0.006$)&  $107.3 \pm 2.7$ &$ 51.7\pm1.6$\\
 CuO & $2.162\pm 0.029$ ($0.748\pm0.062$)&$1.699\pm 0.001$ ($0.189\pm 0.004$)&$2.042\pm 0.003$ ($0.392\pm0.007$)&  $145.8 \pm 1.5$ &\\
\hline
\end{tabular}
\caption{Energies (and FWHM) of the transitions to the $d_{3z^2-r^2}$, $d_{xy}$, $d_{xz/yz}$ states (in eV) extracted from Cu L$_3$ edge RIXS spectra and energies of magnetic excitations $\Delta E$ (in meV) extracted from Cu and Ni L$_3$ edge RIXS spectra taken at 30\,K and $2\theta=90^\circ{}$ as explained in the text. }\label{tab:dd_J}
\end{table*}

The peak corresponding to the $d_{3z^2-r^2}$ excited state is centered at approximately 0.9\,eV. This value is much lower than the least distorted layered cuprate (1.7\,eV in \ch{La_2CuO_4} \cite{sala2011}), still it implies that the symmetry of the Cu sites is lower than octahedral and points at a sizable distortion of the CuO$_6$ octahedron in all samples, though with appreciable differences (see Tab.\,\ref{tab:dd_J}). Namely, the energy of the $d_{3z^2-r^2}$ excited state and its full width at half maximum (FWHM) in the noNi sample is higher than in the other compounds, indicating that the distortion of the Cu sites is larger and more broadly distributed in this sample than in the other compounds, and suggesting that the presence of Ni in the other systems reduces positional disorder. A similar, though less effective, role is played by Mg. These findings are consistent with XRD, PDF and hard XAS investigations. \par

Finally, we look at the changes in the RIXS spectra as a function of temperature. A similar behavior is observed for all compounds, so we report in Fig.\,\ref{fig:Cuedge}\,(c) only the data for 5HEO as an example. We note that $d$-$d$ excitations soften and broaden in energy as the temperature is raised; the softening can be attributed to thermal expansion, as an increase in the Cu-O distance weakens the crystal-field interaction, while the broadening is likely caused by lattice vibrations, which induce fluctuations in the bond length, hence a distribution of Cu sites with varying crystal-field interactions, a phenomenon previously observed in NiO \cite{ishikawa2017} and CuO \cite{huotari2014}.

\begin{figure*}
    \centering
   \includegraphics[width=0.75\textwidth]{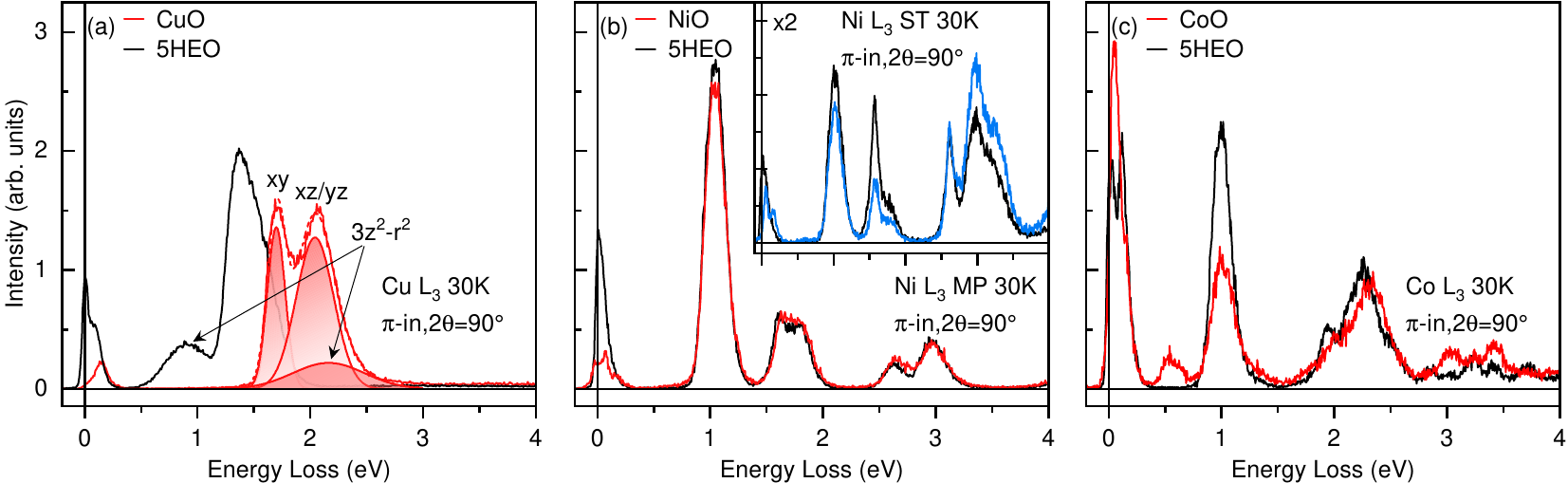}
    \caption{(a) Comparison between the Cu (a), Ni (b) Co (c) L$_3$ edge RIXS spectra of 5HEO and the single binary oxides, CuO NiO and CoO, respectively, taken at 30\,K. The inset of panel (b) shows the RIXS spectrum acquired at satellite peak in the XAS profile, as described below in the main text.}
       \label{fig:heo_Cu_Ni_Co}
\end{figure*}

Our experimental findings provide unambiguous evidence that the Cu sites are affected by a sizable Jahn-Teller distortion, which breaks the cubic symmetry of the \ch{CuO_6} octahedron, and back up previous measurements using structural probes \cite{berardan2017, rost2017, fracchia2020, sushil2021}. While we cannot estimate the variation of the Cu-O bond lengths with respect to the ideal octahedral situation, we can estimate the energy change involved in the distortion by comparing the RIXS spectra of rocksalt 5HEO and tenorite CuO, as shown in Fig.\,\ref{fig:heo_Cu_Ni_Co}\,(a). The spectra are remarkably different and, specifically, the average energy of $d$-$d$ excitations is significantly lower in 5HEO than in CuO, consistent with a larger Cu-O bond length. The fit to the CuO spectrum following the same procedure described above puts the $d_{3z^2-r^2}$ excited state at an energy of approximately 2.2\,eV, in perfect agreement with previous estimates \cite{huotari2014}, i.e., approximately 1.3\,eV higher than in 5HEO. 
It should be possible, though beyond the scope of this work, to extract a more accurate estimate of the enthalpy of mixing from the observed reconstruction of the electronic structure.

\begin{figure*}
    \centering
   \includegraphics[width=0.75\textwidth]{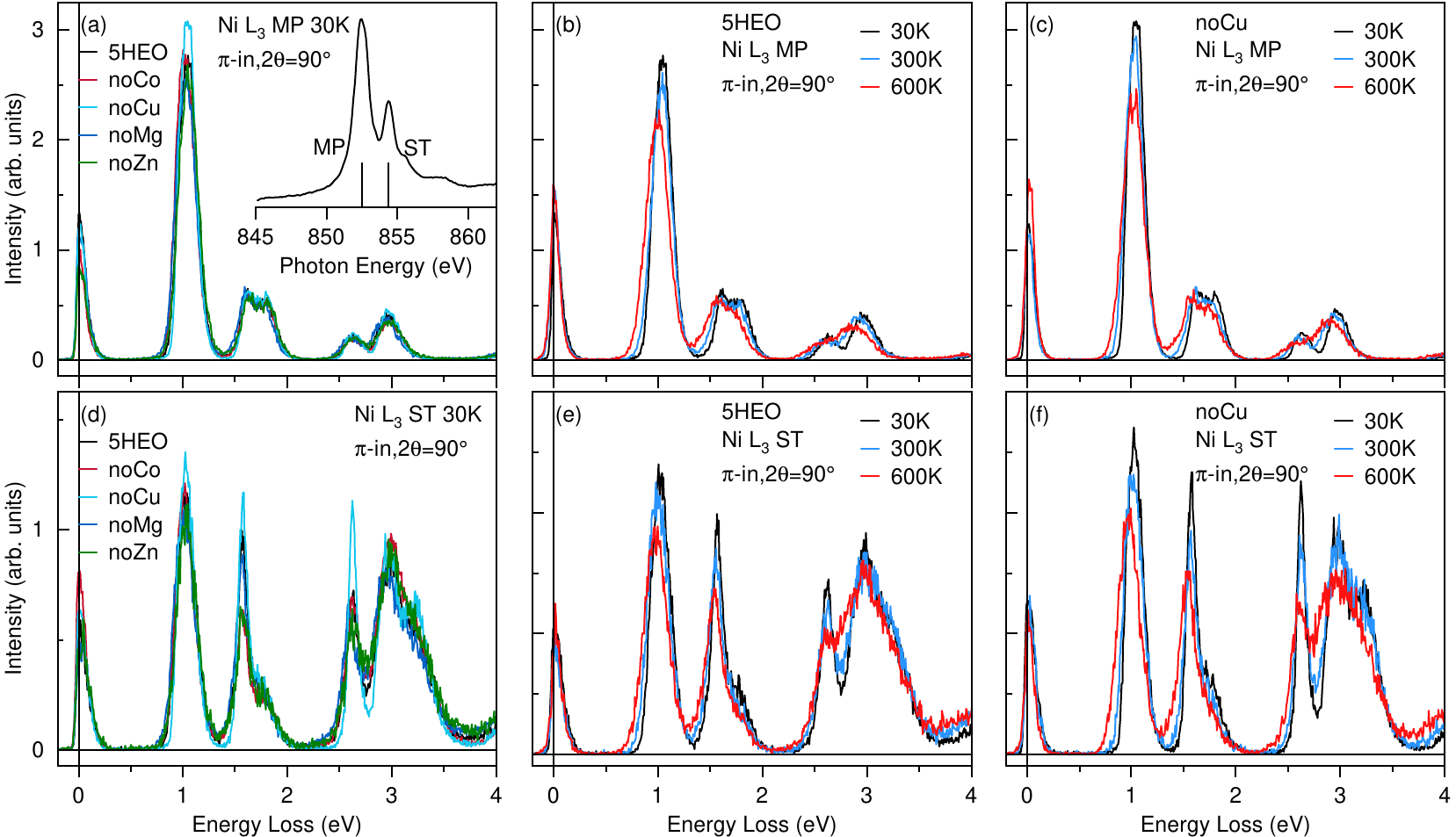}
    \caption{(a) \ch{Ni} L$_3$ edge RIXS spectra taken at 30\,K with an energy of the incident photons corresponding to the main peak (MP) in the XAS profile (vertical red bar in the inset). Temperature dependence of the RIXS spectra of 5HEO (b) and (c) noCu. (d)-(f) Same as (a)-(c), but with an energy of the incident photons corresponding to the satellite peak (ST) in the XAS profile (vertical blue bar in the inset of panel (a)).}
       \label{fig:Niedge}
\end{figure*}

\begin{figure*}
    \centering
   \includegraphics[width=0.75\textwidth]{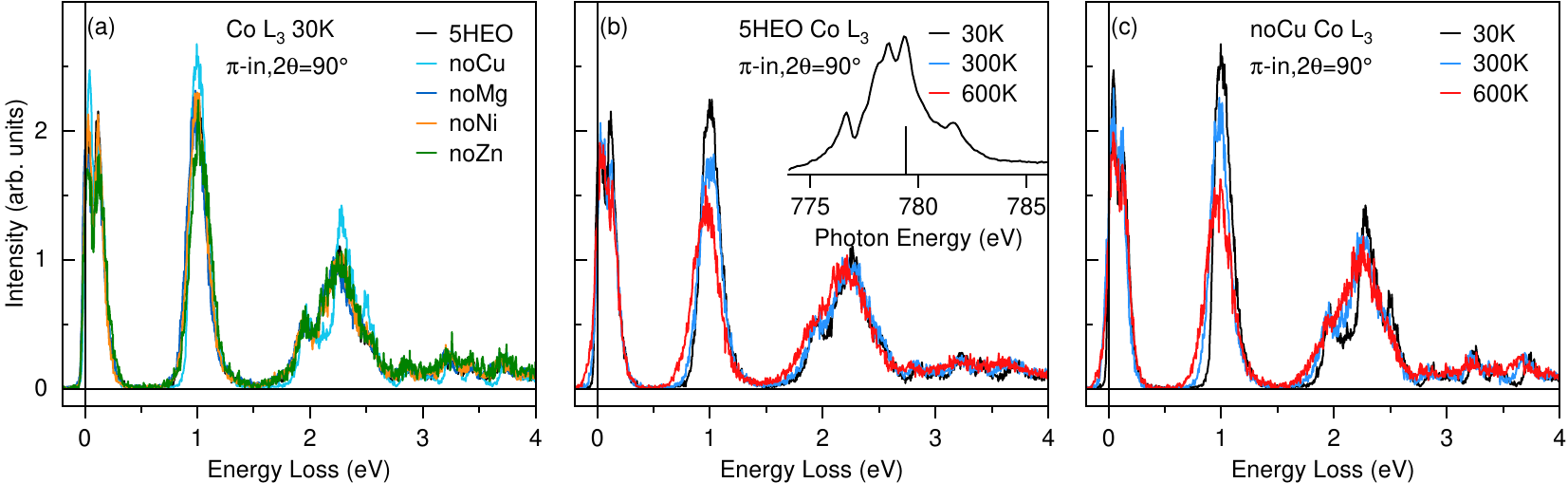}
    \caption{(a) \ch{Co} L$_3$ edge RIXS spectra taken at 30\,K with an energy of the incident photons corresponding to the main peak in the XAS profile (vertical red bar in the inset). Temperature dependence of the RIXS spectra of 5HEO (b) and (c) noCu.}
       \label{fig:Coedge}
\end{figure*}

Next, we look for signatures of distortions also at the Ni and Co sites. Fig.\,\ref{fig:Niedge} (a) and (d) show the low-temperature \ch{Ni} L$_3$ edge RIXS spectra with energies of the incident photons corresponding to the main (MP) and satellite (ST) peaks in the XAS profile at approximately 852 and 854\,eV, respectively. In analogy to NiO \cite{Ghiringhelli_NiO_PRL, Betto_PRB, Nag_PRB}, the energy range from 1 to 3.5\,eV is covered by $d$-$d$ excitations, while charge-transfer excitations emerge above 3.5\,eV; single and double spin excitations appear below 0.1\,eV. Notably, the RIXS spectra at the MP look very much alike for all samples. On the contrary, the RIXS spectra excited at the ST peak show differences; in particular, we observe that the RIXS spectrum of noCu is significantly sharper than the spectra of the other compounds. Fig.\,\ref{fig:Coedge}\,(a), instead, shows the low-temperature \ch{Co} L$_3$ edge RIXS spectra with an energy of the incident photons corresponding to the main peak of the XAS profile at 779\,eV. Here, $d$-$d$ and charge-transfer excitations are dominant down to the lowest energies because of the multiplet structure of \ch{Co}$^{2+}$ \cite{Shi_PRB,Magnuson2002,tomiyasu2006}. As a result, if present, magnetic excitations cannot be easily identified. Similarly to the \ch{Ni}, also the Co L$_3$ edge RIXS spectra overlap for all the samples, with the exception of noCu, which again shows sharper features.

Both Ni (see Fig.\,\ref{fig:Niedge}\,(b), (c), (e) and (f)) and Co (see Fig.\,\ref{fig:Coedge} (b) and (c)) L$_3$ edge RIXS spectra of 5HEO and noCu broaden with temperature, progressively eliminating their differences. Consequently, there appears to be a potential parallel between the broadening of the RIXS spectra induced by temperature and the incorporation of Cu in the crystal structure. Since the FWHM of the peaks reflects a distribution of bond lengths \cite{ishikawa2017}, we conclude that the bond length in noCu is more uniformly distributed than in the other compounds. Unfortunately, we cannot rule out the possibility that noCu itself exhibits distortions, nor can we quantify the crystal-field splitting associated with the distortions at transition metal sites other than Cu. These remain as subjects of research for future investigations.

Similarly to the case of CuO, we try to highlight potential variations in the electronic structure of rocksalt 5HEO compared to that of pristine rocksalt NiO and CoO in Fig.\,\ref{fig:heo_Cu_Ni_Co}\,(b) and (c), respectively. Unlike the case of CuO, the spectra of NiO are not markedly different from those of 5HEO, the major changes being related to the intensity, not the energy, of the various excited states. The comparison between 5HEO and CoO, instead, shows little, but appreciable differences in the energy of $d$-$d$ excitations, especially in the very low energy region, where the double structure in 5HEO merges into a single peak in CoO at approximately 0.1\,eV (note that the excitation at approximately 0.5\,eV is related to \ch{Co^{3+}} impurities). This suggests that the electronic structure of \ch{Ni^{2+}} ions does not change significantly when NiO is incorporated in 5HEO, implying a negligible enthalpy of mixing, while the electronic reconstruction of \ch{Co^{2+}} ions is appreciable. Unfortunately, we could not perform the same comparison with MgO and ZnO because $d$-$d$ excitations should not occur in systems with nominal $d^0$ and $d^{10}$ electronic configuration, respectively.

\subsection{\label{subsec:Cu L3_maggnon}Magnetic excitations and interactions}

\begin{figure*}
    \centering
    \includegraphics[width=0.75\textwidth]{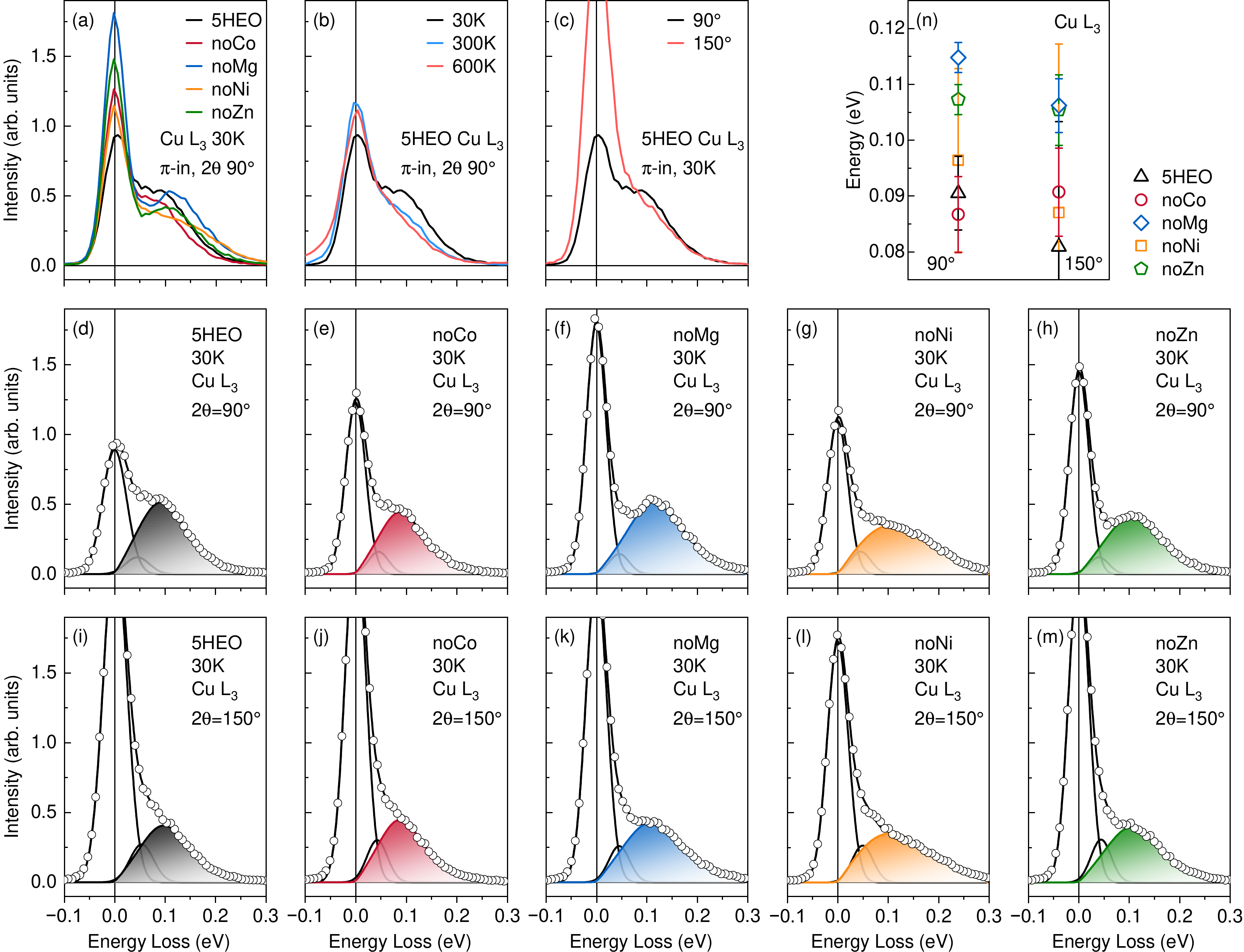}
    \caption{(a) Comparison of the low-energy region of the Cu $L_3$ edge RIXS spectra taken at 30\,K and 2$\theta=90^\circ{}$. Temperature (b) and 2$\theta$ (c) dependence of the RIXS spectra of 5HEO. Experimental and fitting curves of the RIXS spectra taken at 30\,K of all samples for 2$\theta = 90^\circ{}$ (d)-(h) and $150^\circ{}$ (i)-(m). (n) Spin excitation energy as extracted from the fitting for 2$\theta = 90^\circ{}$ (left) and $150^\circ{}$ (right).
}
       \label{fig:RIXS_magnonCu}
\end{figure*}

\begin{figure*}
    \centering
    \includegraphics[width=0.75\textwidth]{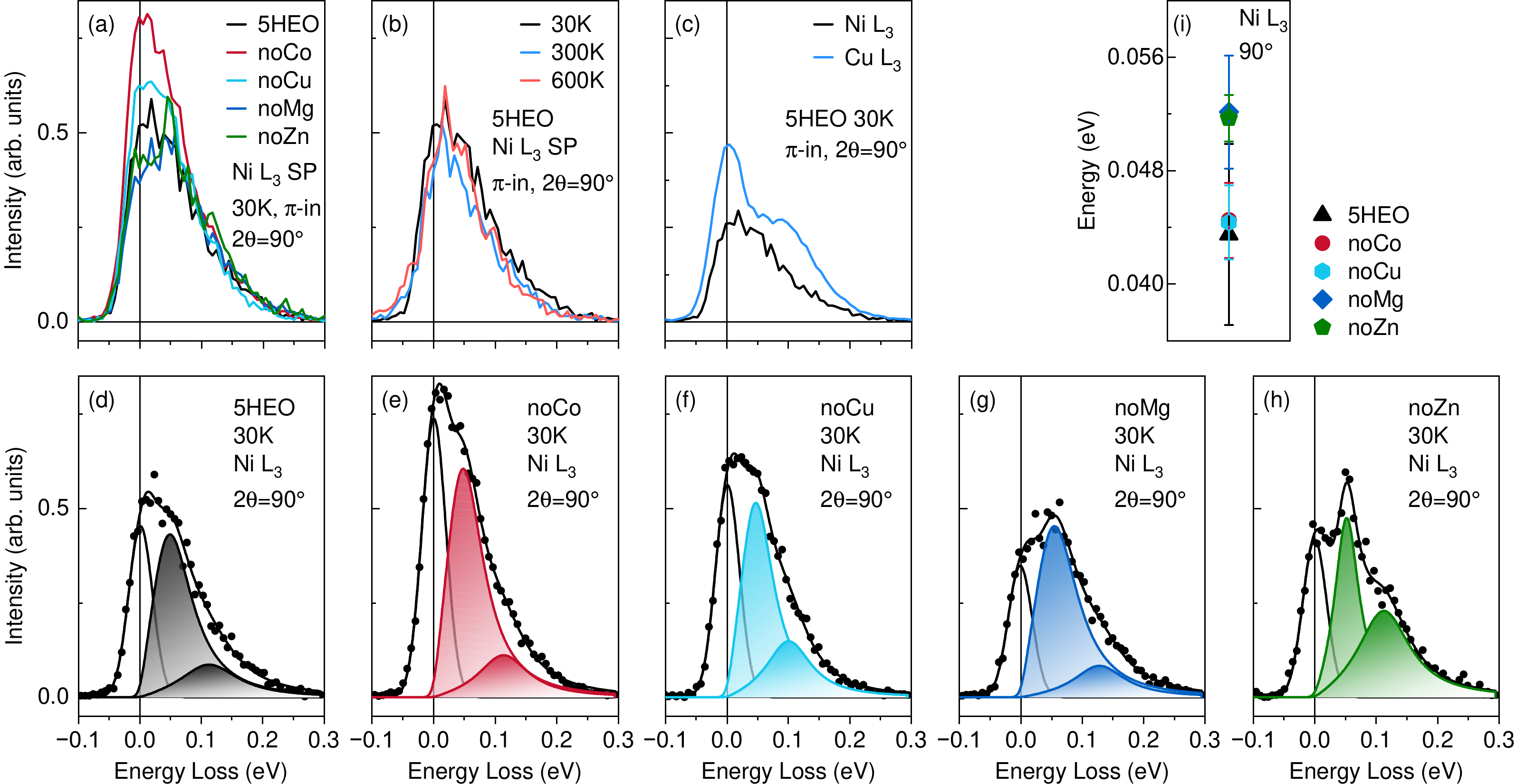}
    \caption{(a) Comparison of the low-energy region of the Ni $L_3$ edge RIXS spectra taken 30\,K and 2$\theta = 90^\circ{}$. (b) Temperature dependence of the RIXS spectra of 5HEO. (c) Comparison of the Cu and Ni L$_3$ edge RIXS spectra of 5HEO. (d)-(h) Experimental and fitting curves of the RIXS spectra taken at 30\,K of all samples. (i) Spin excitation energy as extracted from the fitting.
}
       \label{fig:RIXS_magnonNi}
\end{figure*}

We now focus on the low-energy region of the RIXS spectra to investigate the magnetic dynamics in 5HEO and related compounds. Fig.\,\ref{fig:RIXS_magnonCu}\,(a) compares the Cu L$_3$ edge RIXS spectra of all samples at 30\,K, where the feature at 0.1\,eV is better appreciated. The magnetic nature of this excitation can be attributed by drawing an analogy to cuprates \cite{Braicovich_PRL2010}, and here confirmed by two additional experimental observations: i) Fig.\,\ref{fig:RIXS_magnonCu} (b) shows that its intensity progressively decreases with increasing temperature from below to well above the Néel temperature of 5HEO; ii) the RIXS spectra were fit in Fig.\,\ref{fig:RIXS_magnonCu}\,(d)-(h) to two energy-resolution-limited Gaussian curves (black solid lines) representing the quasi-elastic and phonon peaks and a Pearson curve (colored solid line with shaded area) representing the magnetic excitation. The results are summarized in the left panel of Fig.\,\ref{fig:RIXS_magnonCu}\,(n) and in Tab.\,\ref{tab:dd_J} and show that the energy of the excitation is consistently higher in noMg and noZn than in the other compounds; since Mg$^{2+}$ and Zn$^{2+}$ are non magnetic, the effect can be qualitatively accounted for by noting that their removal from the crystal structure increases the energy cost of flipping a Cu spin, consistent with the magnetic origin of the feature. This argument is better developed below to provide further information. Also for magnetic excitations, the peak is broader in the noNi sample than in the other compounds, again suggesting that \ch{Ni^{2+}} reduces positional disorder in the system.

Having confirmed that the origin of the low-energy excitation is magnetic, we discuss its local or collective nature. In single crystalline materials this is usually tested by checking the dispersion of magnetic excitation, but for powder samples the test could be less conclusive. In our case, however, powder averaging is not very effective because of the small momentum available in the soft X-ray energy range and the experiment might still be meaningful. With this in mind, we repeated the RIXS measurements for 2$\theta=150^\circ{}$, where 2$\theta$ is related to the modulus of the momentum transfer by $|\mathbf{q}|=4\pi/\lambda \sin(2\theta/2)$ and $\lambda=13.3$\,\AA\ is the wavelength of the incident and scattered photons. The spectra are shown in Fig.\,\ref{fig:RIXS_magnonCu} (i)-(m) and the results of the fitting summarized in the right panel of Fig.\,\ref{fig:RIXS_magnonCu}\,(n). Within the experimental uncertainties, we find that the energy of the magnetic excitations does not depend on $|\mathbf{q}|$, so we conclude that they should be better thought of as local spin excitations rather than collective spin-waves. This result seems reasonable for a system with large configurational disorder, which prevents the magnetic excitations from propagating coherently over long distances. 

We identify magnetic excitations also in \ch{Ni} L$_3$ edge RIXS spectra taken at the ST peak (see Fig.\,\ref{fig:RIXS_magnonNi}), in analogy to NiO \cite{Ghiringhelli_NiO_PRL, Betto_PRB, Nag_PRB}. Their intensity is strong at low temperature (Fig.\,\ref{fig:RIXS_magnonNi}\,(b)). As shown in Fig.\,\ref{fig:RIXS_magnonNi} (d)-(h), the RIXS spectra were fit to a resolution-limited Gaussian curve  (black solid line) representing the quasi-elastic peak and two antisymmetrized Lorentzian curves (colored solid line with shaded area) representing the single and double magnetic excitations \cite{Nag_PRB}. Fig.\,\ref{fig:RIXS_magnonNi} (i) and Tab.\,\ref{tab:dd_J} report the energy of the single spin excitation. In Fig.\,\ref{fig:RIXS_magnonNi}\,(c) we directly compare the Cu and Ni L$_3$ edge RIXS spectra of 5HEO and observe that the energy of the single magnetic excitation measured at the \ch{Ni} resonance is significantly lower than that measured at the \ch{Cu} resonance. In our view, this can be explained as in the following: configurational disorder prevents excitations from developing long-range coherence, while the chemical selectivity of RIXS probes only those localized at the resonant metal ions, characterized by different magnetic interactions each.

Based on the considerations above, we use a cluster composed by a central metal ion \ch{M} coordinated to six (next-nearest neighboring) metal ions through $180^\circ{}$ bonds, and adopt an Ising-like Hamiltonian to model its magnetism and explain the dependence of the spin excitation energy on the molar fraction of the various elements as probed by RIXS at the various absorption edges. We introduce the spins of the central and of the surrounding metal ions, $\mathrm{S}_\mathrm{M}$ and $\mathrm{S}_j$, respectively, and their magnetic (super-exchange) interactions $J_{\mathrm{M},j}$. We neglect the magnetic coupling of the central metal ion to its twelve nearest neighbor metal ions through $90^\circ{}$ bonds, as it is expected to be very weak \cite{rak2020}; then, the energy cost associated to a  $\Delta S_\mathrm{M}=1$ spin excitation of the central metal ion in the magnetically ordered phase is given by
\begin{equation}
     \Delta E^\mathrm{M}=\sum_{j=1}^6 J_{\mathrm{M},j}S_j = 6\sum_i\chi_i J_{\mathrm{M},i} \mathrm{S}_i ,
\end{equation}
where the second equality follows from taking into account that the each of the six sites surrounding the central metal ion is occupied by the $i$-th element with probability $\chi_i$, the corresponding molar fraction. Since $\mathrm{S}_i=0$, 3/2, 1, 1/2 and 0 for $i=\ch{Mg}$, \ch{Co}, \ch{Ni}, \ch{Cu} and \ch{Zn}, respectively, it turns out that
\begin{equation}\label{eq:energy spin flip}
     \Delta E^\mathrm{M} = 9\chi_\mathrm{Co}J_{\mathrm{M},\mathrm{Co}}+6\chi_\mathrm{Ni}J_{\mathrm{M},\mathrm{Ni}}+3\chi_\mathrm{Cu}J_{\mathrm{M},\mathrm{Cu}},
\end{equation}
where the molar fractions of the various metal ions depend on the actual composition of the sample.

Such a simple model captures two key features of our experimental observations: it explains the dependence of the spin excitation energy on i) the chemical composition of the various samples and ii) the absorption edge used for RIXS measurements, which selects the central (resonant) metal ion. Specifically, the spin excitation energy in samples with a non magnetic ion removed, i.e., noMg and noZn, is larger than in 5HEO by a factor 5/4=1.25 according to Eq.\,\ref{eq:energy spin flip}, irrespective of the central (resonant) metal ion. Indeed, inspection of Tab.\,\ref{tab:dd_J} shows that the ratios between the experimental values of the spin excitation energies for noMg (noZn) and 5HEO is 1.27 (1.19) and 1.20 (1.19) at the Cu and Ni edges, respectively. We pushed the analysis of the data using the model even further and used it to estimate the magnetic interactions between pairs of magnetic ions by fitting the calculated spin excitation energies to the values observed in our experiment. We simultaneously fit the results from \ch{Cu} and \ch{Ni} edge RIXS measurements as they are coupled through the magnetic interaction term $J_{\mathrm{Ni},\mathrm{Cu}}=J_{\mathrm{Cu},\mathrm{Ni}}$ and obtain the values of the magnetic interactions reported in Tab.\,\ref{tab:J_values}. We note that the values of the magnetic interactions are not necessarily consistent with the theoretical predictions \cite{rak2020}, but are compatible with available experimental estimates for related systems (e.g., \ch{La_2CuO_4}\cite{peng2017_Nature} and \ch{NiO}\cite{Betto_PRB}) with similar bond geometries. Discrepancies, instead, can possibly be attributed to different bond lengths, although we find a larger $J_{\mathrm{Ni},\mathrm{Ni}}$ magnetic coupling despite the larger Ni-O-Ni distance in 5HEO (4.23\,\AA) than in NiO (4.17\,\AA). Unfortunately, we did not find estimates of the magnetic interactions between other pairs of magnetic ions, for which RIXS or inelastic neutron scattering studies of oxides like \ch{La_2Cu_{1-x}Co_xO_4} \cite{dash2024}, \ch{La_2Ni_{1-x}Co_xO_4} \cite{amow2004} and \ch{La_2Cu_{1–x}Ni_xO_4} \cite{Ting_PRB} would be particularly useful.

\begin{table}[]
\renewcommand{\arraystretch}{1.5}
\centering
\caption{Values of the magnetic interactions (in meV) extracted from the RIXS data as explained in the text and compared with theoretical values for 5HEO from Ref.\cite{rak2020} and previous works on related systems.}\label{tab:J_values}
\begin{tabular}{cccc}
 \hline
 &  This work  & Theory \cite{rak2020} & Previous works \\ 
 \hline
 $J_{\mathrm{Cu},\mathrm{Cu}}$  &96.5 $\pm$ 12.1  &37.0 &  140.0 (\ch{La_2CuO_4} \cite{peng2017_Nature})
 \\
 $J_{\mathrm{Cu},\mathrm{Co}}$  &9.3 $\pm$ 2.9  &18.2&  \\
 $J_{\mathrm{Cu},\mathrm{Ni}}$  &11.6 $\pm$ 3.8  &21.7&   \\
 $J_{\mathrm{Ni},\mathrm{Co}}$  &3.4 $\pm$ 1.3  &12.8&  \\
 $J_{\mathrm{Ni},\mathrm{Ni}}$  &23.9 $\pm$ 2.6  &18.9&  19.0 (\ch{NiO} \cite{Hutchings1972})  \\  
$J_{\mathrm{Co},\mathrm{Co}}$  &  &10.6  & 2.4 (\ch{CoO} \cite{Sarte_PRB_2018,Sarte_PRB_2019} )\\
 \hline
\end{tabular}
\end{table}

For the sake of completeness, we predict the spin excitation energies that could potentially be measured by means of \ch{Co} L$_3$ edge RIXS if one were capable of isolating the magnetic signal from the low-energy $d$-$d$ excitations (e.g., by improving the energy resolution or analyzing the polarization of the scattered beam). All the magnetic interactions have been determined in this work, except for $J_{\mathrm{Co},\mathrm{Co}}$; using the literature value of 2.4\,meV for \ch{CoO} \cite{Sarte_PRB_2018,Sarte_PRB_2019} we predict  $\Delta E^\mathrm{Co} \approx 13.6$\,meV for 5HEO, 17.0\,meV for noMg and noZn, 12.0\,meV for noNi and 10.5\,meV for noCu.

\section*{\label{sec:Conclusions}Conclusions}

We conducted a thorough investigation of the structural, electronic, and magnetic properties of 5HEO and related compounds. We probed their long-range crystal structure by means of XRD and concluded that noCu is perfectly cubic, noNi and noMg are tetragonal; noZn and noCo, while remaining cubic, show signs of considerable long-range disorder. The short-range crystal structure has been probed by PDF, XANES and EXAFS on 5HEO, noCu and noMg, where the latter two have been chosen as
references for cubic and tetragonal crystal structures, respectively. It turns out that noMg has the largest short-range disorder, with Ni accommodating the distortion induced by Jahn-Teller active Cu$^{2+}$ ions, and the Co-O bond length being very little affected.

Since all quaternary samples contain 25\% of Cu$^{2+}$, Cu content alone does not account for the emergence of tetragonal symmetry in noMg and noNi. To assess the role of the various chemical species further, we employed RIXS. RIXS measurements at the Cu L$_3$ edge revealed Jahn-Teller distortions at the Cu sites through the splitting of $d$-$d$ excitations. Also, their broadening confirms a high degree of short-range disorder in noNi and noMg, suggesting that Ni$^{2+}$ and Mg$^{2+}$ mitigate positional disorder. Consistently, RIXS spectra at the Co and Ni L$_3$ edges suggest that Cu$^{2+}$ promotes positional disorder. In addition to that of Ni, we could here appreciate an effect on the electronic structure of Co, which was invisible to the analysis of EXAFS data.

Finally, we identified low-energy features in Cu and Ni L$_3$ edges RIXS spectra, which we interpreted as local magnetic excitations. We estimated most of the magnetic interactions between pairs of magnetic Cu$^{2+}$, Ni$^{2+}$ and Co$^{2+}$ ions, the most relevant ingredients for a comprehensive understanding of the magnetic properties of 5HEO. Our findings provide valuable insights into the influence of the various chemical species in the formation and on the magnetic properties of 5HEO and related HEO systems.

\section*{Author contributions}
M.F., M.C., P.G. and M.M.S. conceived the experiment. M.Z., M.F., M.C., G.M. and M.M.S.  analyzed the data. M.F. and G.M. synthesized the samples. M.Z., M.F., G.M., D.I., N.B., L.G., K.K., F.R., M.A. and M.M.S. performed the experiments. M.F., M.C., G.G., P.G. and M.M.S.  supervised the project. All authors contributed to the interpretation. M.Z. and M.M.S. wrote the manuscript with input from all authors.

\section*{Conflicts of interest}
There are no conflicts to declare.

\section*{Data availability}
RIXS data are available on: https://doi.org/10.15151/ESRF-ES-1335198610. XRD, XANES, EXAFS and PDF data are available on https://doi.org/10.5281/zenodo.15784409.\par

\section*{\label{sec:Acknowledgments}Acknowledgments}
We acknowledge ESRF - The European Synchrotron for provision of synchrotron radiation facilities under proposal numbers HC-6743 and HC-5356 and we would like to thank the whole technical and scientific staff of ID32 for their excellent support. The work here presented is partly funded by the European Union – Next Generation EU - ``PNRR - M4C2, investimento 1.1 - Fondo PRIN 2022'' - ``Superlattices of relativistic oxides''  (ID 2022L28H97, CUP D53D23002260006).
The authors acknowledge the ESRF also for provision of beamtime on ID15A (exp CH-5675), thanking Dr. Stefano Checchia for support. They also acknowledge the Elettra synchrotron facility for provision of beamtime (beamline XAFS, experiment 20212199) thanking Dr. Luca Olivi for the kind support and the Diamond Light Source for the provision of in-house beamtime (beamline I20). M. F., M. C., G.M. and P. G. acknowledge support from the Ministero dell’Università e della Ricerca (MUR) and the University of Pavia through the program ``Dipartimenti di Eccellenza 2023–2027''.




\balance


\bibliography{rsc} 
\bibliographystyle{rsc} 
\end{document}